\documentclass{camera}
\usepackage{graphicx}  % uncomment this if your want to insert figures

\def\gsim{\mathrel{\hbox{\rlap{\lower.55ex \hbox {$\sim$}}
           \kern-.3em \raise.4ex \hbox{$>$}}}}
\def\lesssim{\mathrel{\hbox{\rlap{\lower.55ex \hbox {$\sim$}}
           \kern-.3em \raise.4ex \hbox{$<$}}}}

\newcommand{\beq}{\begin{equation}}
\newcommand{\eeq}{\end{equation}}
\newcommand{\ba}{\begin{array}}
\newcommand{\ea}{\end{array}}

\def \etal{{\it et al.~}}
\def \apj{{ \emph {Astrophys. J.}}}
\def \apjl{{ \emph {Astrophys. J. Lett.}}}
\def \apjs{{ \emph {Astrophys. J. Supp.}}}

\begin{document}

%%%%%%%%%%%%%%%%%%%%%%%%%%%%%%%%%%%%%%%%%%%%%%%%%%%%%%%%
% The title, only the first letter capitalized; if you want to split it in
% two or more lines, put a \\ macro at each line break
% example: 
%   \title{Title: first line\\ second line}
%
\title{Observations, theory and implications of thermal emission from gamma-ray bursts} 

%%%%%%%%%%%%%%%%%%%%%%%%%%%%%%%%%%%%%%%%%%%%%%%%%%%%%%%%
% The author(s), separated by commas; do not put a
% comma before the last author, use instead the \and
% macro which produces a normal ``and'' in the
% caps/small caps context
%
\author{A. Pe'er$^1$ \and F. Ryde$^2$}

%%%%%%%%%%%%%%%%%%%%%%%%%%%%%%%%%%%%%%%%%%%%%%%%%%%%%%%%
%
\organization{$^1$ Space Telescope Science Institute, Baltimore, MD
  21218, USA; Riccardo Giacconi Fellow. \\ $^2$ Department of Physics,
  Royal Institute of Technology, AlbaNova, SE-106~91 Stockholm,Sweden}

\maketitle

\begin{abstract}
Recent analyses show evidence for a thermal emission component that
accompanies the non-thermal emission during the prompt phase of GRBs.
First, we show the evidence for the existence of this component;
Second, we show that this component is naturally explained by
considering emission from the photosphere, taking into account high
latitude emission from optically thick relativistically expanding
plasma. We show that the thermal flux is expected to decay at
late times as $F_{\rm BB} \sim t^{-2}$, and the observed temperature as $T \sim
t^{-\alpha}$, with $\alpha \approx 1/2 - 2/3$. These theoretical predictions
are in very good agreement with the observations.  Finally, we
discuss three implications of this interpretation: (a) The relation
between thermal emission and high energy, non-thermal spectra observed
by {\it Fermi}.  (b) We show how thermal emission can be used to
directly measure the Lorentz factor of the flow and the initial radius
of the jet.  (c) We show how the lack of detection of the thermal
component can be used to constrain the composition of GRB jets.

\end{abstract}

%%%%%%%%%%%%%%%%%%%%%%%%%%%%%%%%%%%%%%%%%%%%%%%%%%%%%%%%
% Write the text starting from here and using the usual
% LaTeX commands.
%
\section{Introduction}
\label{sec:intro}

Despite many efforts, a clear understanding of the physical origin of
the photons observed during the prompt emission phase in GRBs is
still lacking. The prompt emission spectra are often fitted with a
broken power law model (known as the ``Band'' function,
\cite{Band93}). However, recent {\it Fermi}-~LAT results show that
this fit is inadequate in some cases in which high energy photons are
observed (e.g., \cite{Abdo+09}). Even more importantly, the ``Band''
function fit does not, by itself, carry any explanation about the
physical origin of the observed photons. A common interpretation is
that the peak observed at sub-Mev range is due to synchrotron emission
\cite{Frontera00}. However, while the synchrotron interpretation is
consistent with many afterglow observations, this interpretation is
inconsistent with the majority of the prompt spectra, due to steep low
energy spectral slopes observed \cite{Preece98}.

This inconsistency motivated us to search for an alternative
explanation. Arguably, the most natural ingredient is a thermal
component, that should exist in the outflow. In principle, thermal
photons can originate either by the initial explosion, or by any
dissipation of the kinetic energy that occurs deep enough in the flow,
in region where the optical depth is $\gg 1$, so that the emitted
photons thermalize before escaping the plasma. 

In a series of papers \cite{Ryde04, Ryde05, PMR05, PMR06, PRWMR07,
  Peer08, RP09, ZP09, Ryde+09} we have extensively studied the
contribution of a thermal emission component to the overall
(non-thermal) GRB prompt spectra. Our research is focused on both the
observational properties \cite{Ryde04, Ryde05, RP09, Ryde+09},
theoretical modeling \cite{PMR05, PMR06, Peer08} and implications of
the existence of this component \cite{PRWMR07, ZP09}. We give here a
brief summary of our key results.

\section{Observational clues}
\label{sec:obs}

Two main difficulties exist in interpreting the observed spectrum. The
first is that clearly, in addition to the thermal component there is a
strong non-thermal part. The second is that the properties of the
thermal component, (temperature and flux) may be time-dependent, hence
its signal is smeared in a time-integrated analysis, as is frequently
done.  In order to overcome the second problem, one needs to carry a
{\it time-dependent} analysis. Such an analysis was indeed carried by
{\it Ryde} (\cite{Ryde04, Ryde05}). In these works, a hybrid model
(thermal + a single broken power law) was found to adequately describe
the prompt spectra of 9 burst over the limited BATSE energy
band. Moreover, the temperature of the thermal component showed a
repetitive behaviour: a broken power law in time, with power law index
$T(t) \propto t^{-2/3}$ after the break time at few seconds.

\begin{figure}
\begin{center}
\includegraphics[width=8.0cm]{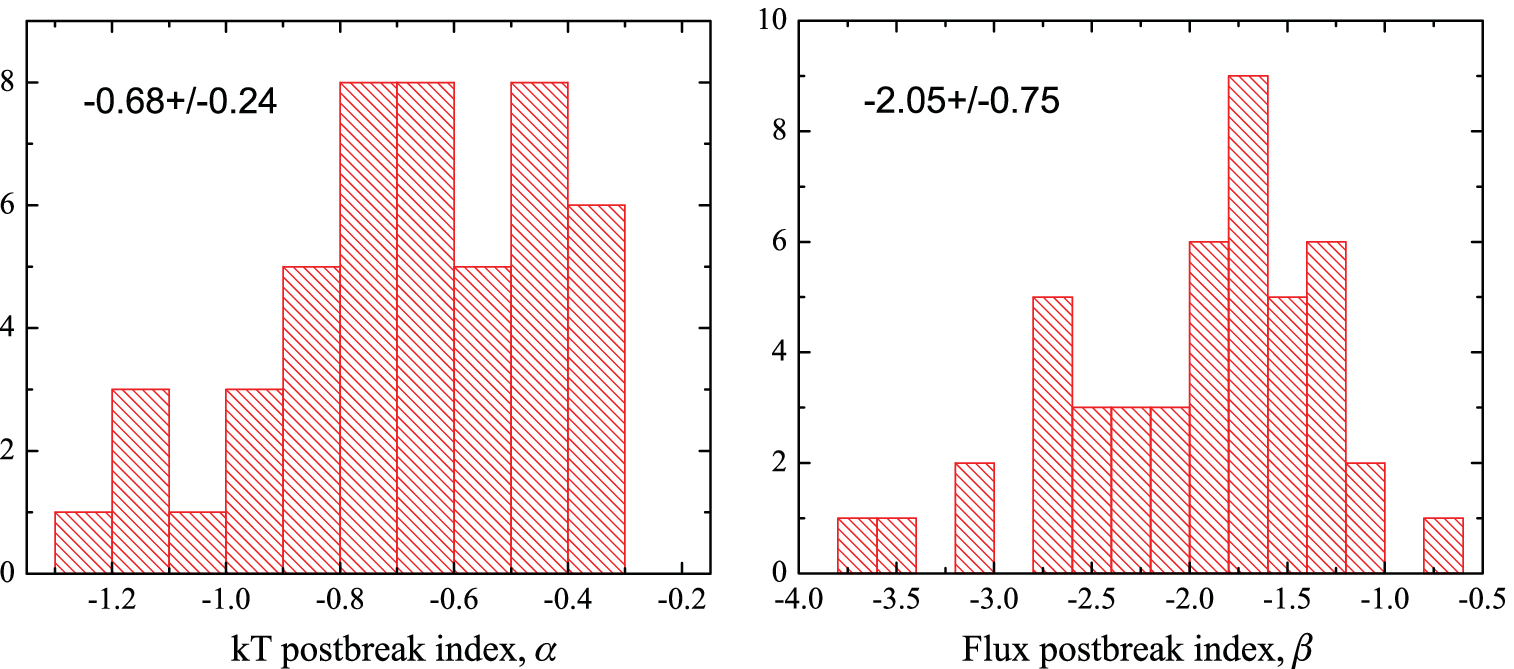}
\includegraphics[width=4.4cm]{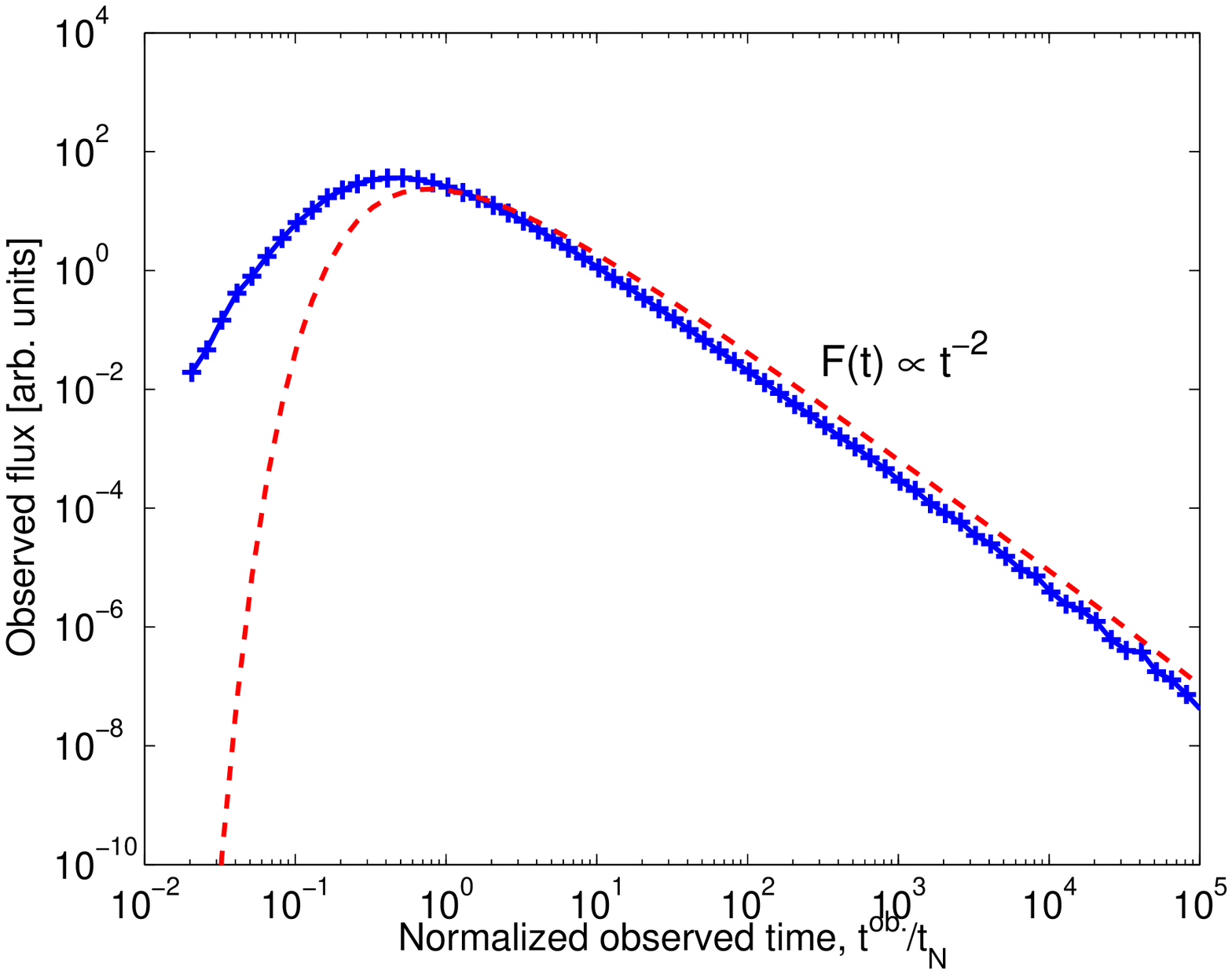}
\end{center}
\caption{(Left, center): Histograms of the late time decay indices of the temperature
  (left) and thermal flux (center) in a sample of 56 bursts (taken from
  \cite{RP09}). Right: theoretical results of the flux decay at late
  times, showing $F(t) \propto t^{-2}$ (taken from \cite{Peer08})}
\label{fig:1} % optional figure label, must be unique
\end{figure}

Repeating a similar analysis on a larger sample of 56 bursts, we found
(\cite{RP09}) that the same repetitive behaviour is ubiquitous.
Moreover, in this work we also considered the evolution of the thermal
flux, and found that it too shows a similar behavior: a broken power
law, with index $F(t) \propto t^{-2}$ at late times. The break time, not
surprisingly, is the same within the errors to the break time found in
the temperature behaviour. Histograms of the late time power law
indices are shown in figure \ref{fig:1}.

% For Figures insertion uncomment the next section

\section{Theoretical interpretation}
\label{sec:theory}

As thermal photons originate from below the photosphere, one needs to
study the properties of the photosphere in a relativistically expanding
plasma. In such a plasma, Lorentz aberration plays a significant role:
for example, the photospheric radius strongly depends on the angle to
the line of sight, $\theta$, via $r_{ph}(\theta) \propto
(\theta^2/3+\Gamma^{-2})$, where $\Gamma$ is the factor Lorentz of the
outflow \cite{Peer08}. This strong dependence implies that photons
emitted from the photosphere at high angles are significantly delayed with
respect to photons emitted on the line of sight: $\Delta t^{ob}
\approx 30 L_{52} \Gamma_2^{-1} \theta_{-1}^4$~s, where $L$ is the GRB
luminosity and $Q_x = Q/10^x$ in cgs units is used.

Calculations of the expected decay of the flux and temperature at late
times are carried under the assumption that the source (the inner
engine) terminates abruptly at a given time. Photons emitted off axis
are delayed (``high latitude'' emission in {\it optically thick}
expanding plasma), resulting in flux decay at late times. Moreover,
due to both weaker Doppler boosting and energy losses to the expanding
plasma at large radii, these photons' observed temperature is also
lower. By integrating over equal arrival time surface in the entire
space, it was shown (\cite{Peer08}) that the flux decays at late times
as $F(t) \propto t^{-2}$ and the temperature as $T(t) \propto
t^{-\alpha}$, with $\alpha \approx 1/2 - 2/3$. The results of the
theoretical calculations of the flux decay (both numerical and
analytic) are shown in figure \ref{fig:1} (right).

Clearly, the theoretical predictions are in excellent agreement with
the observations. While this by itself does not prove that indeed the
photons that we see are thermal, we find the agreement between theory
and observations, as well as the repetitive behaviour seen, two very
strong, independent indications that we indeed are able to properly
identify the thermal emission component and discriminate it from the
non-thermal part.

\section{Implications}
\label{sec:implications}

The existence of a thermal emission component, which, as described, is
a natural outcome of the fireball model, can bring
a breakthrough in our understanding of the physics of GRB prompt
emission. We discuss here three important implications of it.

{\bf The relation between thermal and non-thermal emission}. As GRB
prompt spectra is non-thermal, clearly, in addition to any thermal
component there is a non-thermal part. Thus, according to our picture,
the observed spectrum is composed of (at least) two separated
ingredients, thermal component originating from the photosphere, and
non-thermal component originating from the kinetic energy dissipation
above the photosphere. The true properties of this non-thermal part
can only be deduced after subtracting the contribution from the
thermal part. This, however, is not an easy task: since thermal
photons serve as seed photons to Compton scattering by energetic
electrons (produced by the dissipation mechanism above the
photosphere), they contribute to the cooling of these
electrons. Hence, the resulting non-thermal spectrum (from, e.g.,
synchrotron emission or Compton scattering) depends not only on the
properties of the acceleration mechanism (e.g., the power law) but
also on the relative contribution of the thermal photons. This can
lead to a variety of very complex spectra (see \cite{PMR05, PMR06}),
which depend on the various parameters - the photospheric radius, the
dissipation radius, magnetic field strength, etc. Interestingly, we
were able to show that for a relatively large parameter space region,
a single power law may be sufficient to model the non-thermal part
over a limited energy range. This result is indeed consistent with the
single-power law fitting of the non-thermal part of the spectrum seen
in several recent {\it Fermi} bursts (e.g., GRB090902B,
\cite{Abdo+09})

{\bf Measuring the parameters of the outflow}. One major advantage of
identifying the thermal emission, is that its radius of origin is known
- it is the photosphere. The ratio of the thermal flux and
temperature, $\mathcal{R} \equiv (F/ \sigma T^4)^{1/2}$ ($\sigma$ is
Stefan's constant) must therefore be proportional to the photospheric
radius (for photons emitted on the line of sight). In fact, due to
Lorentz aberration, we showed (\cite{PRWMR07}) that $\mathcal{R}
\propto r_{ph}(\theta=0)/\Gamma$.

In the classical ``fireball'' model, the photospheric radius depends
only on two parameters: the (kinetic) luminosity, and the Lorentz
factor. Hence, for bursts with known redshift, one can use the three
measurable quantities of emission at the photosphere (thermal flux,
temperature and GRB distance) to deduce the values of the three
unknowns - the luminosity\footnote{Note that there is an uncertainty
  in estimating the kinetic luminosity from the flux. This uncertainty
  can be removed once afterglow measurements are available}, the
Lorentz factor and the photospheric radius itself. Moreover, in the
classical fireball model, the dynamics below the photosphere is fully
determined by the conservation of energy and entropy. One can therefore
use the values of the luminosity and Lorentz factor at the
photospheric radius to deduce the radius at which the initial
acceleration began. This radius is denoted here as $r_0$. Implementing
these ideas, we were able to show that for GRB970828, at redshift
z=0.96, the terminal Lorentz factor is $\Gamma=305 \pm 28$, and $r_0 =
(2.9\pm 1.8)\times 10^8$~cm. The statistical errors in the estimate of
the Lorentz factor, $\pm 10\%$ are by far the smallest from all the
methods known today.

{\bf Deducing the outflow composition}. GRBs show a wide variety of
properties from burst to burst. While in some bursts, thermal emission
component is very pronounced (e.g., in GRB090902B the low energy
spectrum is so steep and the peak is so narrow that it is very
difficult to find an alternative explanation to the peak; see
\cite{Ryde+09}), in other bursts it is much less pronounced. As one
example, in the bright burst GRB080916C, the ``Band'' model provides a
good fit to the spectrum, up to the highest observed photons energies,
at 13.2~GeV \cite{Abdo09}. Opacity arguments can easily show that
these energetic photons cannot originate from the photosphere, but
from some (significantly) larger radius \footnote{Note that opacity
  argument by itself does not give the emission radius; in order to
  obtain that, one needs to specify the relation between the radius
  and the Lorentz factor. The results often used in the literature, $r
  = \Gamma^2 c \delta t$ rely on assumed knowledge of the variability
  time $\delta t$, which is highly uncertain.}, $R_\gamma \gsim
10^{15}$~cm. The lack of the detection of a thermal component as
predicted by the baryonic models strongly suggests that a significant
fraction of the outflow energy is initially not in the ``fireball''
form. Thus, we found \cite{ZP09} the most plausible alternative to be
Poynting flux entrained with the baryonic matter. The ratio between
the Poynting and the baryonic flux in this burst is at least $\sim(15
−- 20)$.

\section{Summary}
\label{sec:summary}

The existence of a thermal emission component, which, as described, is
a natural outcome of the fireball model, has a potential to bring
about a breakthrough in our understanding of the physics of GRB prompt
emission. It is therefore the subject of an extensive on-going
research, from all aspects. We stress, that the different physical
environment prevents using the tools developed for studying the
afterglow in the study of
the prompt emission.\\

%%%%%%%%%%%%%%%%%%%%%%%%%%%%%%%%%%%%%%%%%%%%%%%%%%%%%%%%
% End of the paper
%

{\Large \bf Acknowledgments} \\
We would like to thank Ralph Wijers, Peter M\'esz\'aros, Martin Rees
and Bing Zhang for significant contributions to these works.

\end{document}